\documentstyle[aps,twocolumn,floats,prl]{revtex}

\input epsf

\newcommand{\hmpc}{h^{-1}{\rm\,Mpc}}
\newcommand{\ihmpc}{h{\rm\,Mpc}^{-1}}
\newcommand{\ikms}{{\rm\,(kms^{-1})}^{-1}}

\def\lya{Ly$\alpha$}

\begin{document}
\twocolumn[\hsize\textwidth\columnwidth\hsize\csname
@twocolumnfalse\endcsname
 
\title{Cosmological Limits on the Neutrino Mass from the \lya\ Forest}
\author{Rupert A.C. Croft$^1$, Wayne Hu$^2$ and Romeel Dav\'e$^3$}
\address{${}^1$Astronomy Department, Harvard University, 60 Garden Street,
	Cambridge, MA 02138 \\
        ${}^2$Institute for Advanced Study, School of Natural Sciences, Princeton,
	NJ 08540 \\
        ${}^3$Princeton University Observatory, Princeton, NJ 08544}

\maketitle 
\begin{abstract}

The \lya\ forest in quasar spectra probes scales where massive
neutrinos can strongly suppress the growth of mass fluctuations.
Using hydrodynamic simulations with massive neutrinos,
we successfully test techniques developed to measure the mass power
spectrum
from the forest. A recent observational measurement in conjunction with a 
conservative implementation of other cosmological 
constraints places upper limits 
on the neutrino mass: $m_{\nu} < 5.5$ eV for all values of $\Omega_{m}$, and 
$m_{\nu} \lesssim 2.4 (\Omega_m/0.17 -1)$ eV, if $0.2 \le \Omega_{m} \le 0.5$ as currently observationally
favored (both $95\%$ C.L.).

\end{abstract}
]

Experimental evidence for finite neutrino masses and flavor oscillations
continues to mount.  Recently the Super-Kamiokande experiment has provided
strong evidence that oscillations from $\nu_\mu$ to another species
involve a mass greater than $\sqrt{\delta m^2} =0.07^{+0.02}_{-0.04}$eV 
\cite{SuperK}.  The 
LSND experiment
suggests the existence of 
$\nu_\mu$ to $\nu_e$ oscillations with $\sqrt{\delta m^2} \sim
0.4$ eV \cite{LSND}.  Finally, the solar neutrino deficit requires
$\sqrt{\delta m^2} \sim 0.003$ eV \cite{Solar}.  
These mass splitting results are consistent with one to three weakly interacting
neutrinos in the eV mass range \cite{Sterile}.   

Neutrinos in this mass range are important cosmologically since,
if they exist, they would represent a non-negligible 
contribution to the dark matter
content of the Universe.
In units of the critical density, neutrinos contribute
$\Omega_\nu = N h^{-2} m_\nu/94$ eV, where $h$ is the dimensionless
Hubble constant ($H_0=100h$ km s$^{-1}$ Mpc$^{-1}$), and $N$ is
the number of degenerate mass neutrinos.  As light neutrinos do not 
 cluster on small scales, they retard the
gravitational growth of density fluctuations.
  Any measure of small-scale clustering
is thus sensitive to neutrino masses in this range.  One 
such useful measure is the clustering of the intergalactic medium revealed by
the absorption features in quasar spectra known as the 
\lya\ forest \cite{Lya}. As we will discuss below, the 
\lya\ forest has the distinct advantage that clustering properties of the
mass distribution can be inferred from it, which greatly facilitates
comparisons with theory, and makes 
large searches of parameter-space possible.

In this {\it Letter}, we make use of a recent \lya\ forest
measurement of the power spectrum of mass fluctuations 
\cite{Pettinipk} to place limits on the mass of the neutrino(s). 
Conversion of the power spectrum measurements into neutrino mass limits 
requires a framework for cosmological structure formation.  There is growing 
evidence that structure formed by the gravitational
instability of cold dark matter (CDM) with adiabatic, Gaussian initial 
fluctuations. 
Upcoming Cosmic Microwave
Background (CMB)
experiments should conclusively determine whether this assumption is 
a good one \cite{HuSugSil97}. 
For now, we note that this framework, which we adopt,
includes all currently favored models,
and also that whatever the true model for structure formation, 
\lya\ forest 
measurements should respond sensitively to the presence
of massive neutrinos.
 
Our adiabatic CDM dominated universes are described by 6 free parameters: 
the matter density $\Omega_m$, dimensionless Hubble constant $h$,
baryon density $\Omega_b$,
neutrino density $\Omega_\nu$, density fluctuation
amplitude $A$, and 
tilt $n$, which define the initial density power spectrum 
$P_{\rm init}(k) =A k^n$. We also initially
assume that spatial geometry is flat, as implied by
 recent measurements
of distant supernovae and CMB anisotropies \cite{CMBSN}, 
before investigating
the consequences of relaxing this assumption.

We begin by describing the \lya\ forest power spectrum
measurement method and test it on hydrodynamic simulations.
We then apply the observational 
constraint to the 6 dimensional CDM parameter space
to find an upper limit on the neutrino mass.  
Given this large parameter space, we conservatively employ other
cosmological constraints, notably from the abundance of galaxy clusters and the
age of globular clusters, 
to constrain other parameters
that can mimic the effects 
of massive neutrinos.    
Finally, we consider prospects for making 
a precise measurement of $m_\nu$ using
future \lya\ forest observations and upcoming CMB experiments.  

{\it Testing \lya\ forest simulations with $m_\nu >0$.---}
The \lya\ forest of neutral hydrogen absorption seen in
quasar spectra \cite{Lya}
 arises naturally in cosmological scenarios
where structure forms by the action of gravitational instability.
In hydrodynamic simulations of such models \cite{Hydro} 
 most of the absorption arises in gas of moderate overdensity,
 whose physical state is governed by simple processes
(mainly photoionization heating and adiabiatic cooling, see also
the analytical modeling of \cite{Anal}).
The density field can then be locally related to the optical 
depth for \lya\ absorption \cite{HeliumII}
and hence a directly observable quantity, the transmitted flux in a quasar
 spectrum. 

 The \lya\ forest can therefore be used to determine the statistical
properties of the density distribution, and in particular $P(k)$, 
the power spectrum of density fluctuations. 
A method for carrying this out was described 
 by \cite{Lyapksim}, who also tested it on hydrodynamic 
simulations with CDM only.
The method relies explicitly on the
 assumptions that the initial fluctuations were Gaussian,
 and that gravitational instability was responsible for their growth.
A measurement from an observational dataset was made by \cite{Pettinipk}.
As we will use this result to constrain the neutrino mass,
we first test the method on a hydrodynamic simulation which includes 
massive neutrinos.


The measurement of $P(k)$ 
from \lya\ forest spectra is carried out in two stages. First, the
shape of $P(k)$ is measured from the power spectrum of the \lya\ forest flux.
Second, normalizing simulations are used to set the amplitude
of the linear mass $P(k)$. We refer the reader to \cite{Lyapksim} for details.

The hydrodynamic simulation itself is described in detail by \cite{Lowzlya}. 
We follow the evolution of structure in
a model with two mass-degenerate neutrino species,
using the Parallel TreeSPH hydrodynamic code \cite{PTreeSPH}.
The model parameters are $\Omega_{m}=1, h=0.5, \Omega_{b}=0.075$,
and  $\Omega_{\nu}=0.2$, so that the mass in both species 
combined is $5$ eV.  This so-called cold plus hot dark matter model 
(CHDM) is normalized to fit
the  COBE results \cite{BunWhi97}, so that the amplitude of mass fluctuations
in $8\ \hmpc$ spheres at $z=0$, $\sigma_{8}=0.7$.
We use a box of size $11.111\ \hmpc$, periodic boundary conditions
and initial conditions taken from \cite{CHDMinit}.
The CDM and gas components are represented by $64^{3}$ particles each,
and the neutrinos by $2\times64^{3}$ particles. 
We use the distribution and physical state of the 
gas at $z=2.5$ to generate artificial \lya\ spectra, for 1200 
randomly chosen lines of 
sight through the simulation volume.

We then apply the $P(k)$ recovery method of \cite{Lyapksim}
to these spectra.
We use normalizing simulations run under
the PM approximation \cite{HydroPM,Lyapksim} with 
$64^{3}$ particles and an $11.111 \hmpc$ box.
We use the same estimator for the amplitude of $P(k)$ as in
the observational analysis paper, \cite{Pettinipk}.
 The results of the test are shown in Fig. 1, where we plot the
recovered $P(k)$, together with the linear theory prediction
 for the model.
 We also show the linear power spectrum of  
a CDM-only model (with $\Omega_{m}=1, h=0.5$),
again normalized to COBE, so that $\sigma_{8}=1.2$.

The error bars on the $P(k)$ points are representative of the
``cosmic variance'' error which arises from only having
one hydrodynamic simulation volume. We estimate this uncertainty by
running 10 additional PM approximation simulations of the CHDM model,
and extracting 1200 lines of sight from each. The standard deviation of their 
results for each $P(k)$ point provides the error bar.
 There is an additional 
overall amplitude uncertainty associated with the normalization. We estimate
this by applying the normalizing procedure to
 the PM CHDM simulations taken individually, finding
that the additional error
in this test case is a negligible $4\%$ in $P(k)$.

We use the simulation to  test for systematic errors in the technique.
The observational result was given by \cite{Pettinipk}
in terms of a power law 
fit to the $P(k)$ data points with  $2.7 \times 10^{3} 
\ikms < k < 1.42 \times 10^{2} \ikms$
(which corresponds to $0.5\ \ihmpc < k < 2.7\ \ihmpc$
for an Einstein-de Sitter model).
We fit the simulation data points to a power-law over this range, finding 
an amplitude $\Delta^{2}(k)=0.16\pm0.028$ ($1 \sigma$) 
at $k=1.5 \ihmpc$ [$=0.008 \ikms$]. Here
 $\Delta^{2}(k)=k^{3}P(k)/2\pi^2$,
the contribution to the density field variance from a unit interval
in $\log k$.
 The logarithmic slope, $n=-2.18^{+0.34}_{-0.28}$ ($1 \sigma$).
 The linear theory prediction for CHDM
is $\Delta^{2}(k)=0.21$, $n=-2.40$. Our recovered $P(k)$ is therefore
about $2 \sigma$ too low in amplitude, and has a slightly flatter slope.
 By examining results from
 the more numerous PM simulations, we find that the largest scale
 data point is systematically lowered by peculiar velocity distortions
(as predicted by \cite{Lyavel}), an effect 
which is not accounted for in our estimate of the $P(k)$ shape. Including
or leaving out this point (which has the largest statistical errors) 
has only a small effect on the power law fit.
 It is possible, however, that taking these effects
into account or further refining the analysis
 would improve the result. For the moment it is sufficient to 
note that the observational result of \cite{Pettinipk}
currently has a statistical
uncertainty of $\sim 70\%$ ($2 \sigma$), 
larger than any biases revealed by our test. 

\begin{figure}[t]
\begin{center}
\leavevmode
\epsfxsize=3.4truein \epsfbox{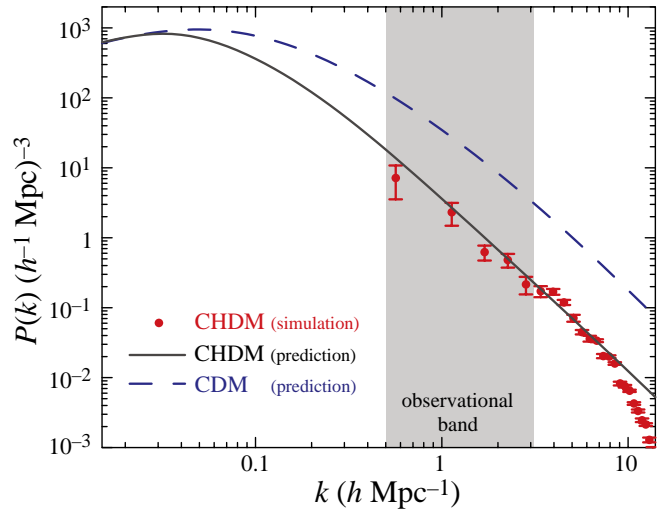}
\end{center}
\caption{
A test of the $P(k)$ recovery method.
The points are $P(k)$ recovered from a hydrodynamic simulation
of the \lya\ forest in a massive neutrino model (CHDM see text).
The turndown for $k> 5 \ihmpc$ is caused mainly by the limited 
resolution of the simulation and lies outside of the observational
band (shaded).
The lines show the linear $P(k)$ predictions for the massive neutrino
model, and for a CDM-only model.
}
\label{fig:neutrino_1}
\end{figure}

{\it Constraining the neutrino mass.---}
Although the suppression of power in the \lya\ forest due to a finite 
neutrino mass is large ($\sim 800 \Omega_\nu/\Omega_m$ percent
\cite{HuEisTeg98}), other aspects
of the cosmological model can counterbalance this effect;
limits on the neutrino mass consequently depend on the range of 
 models allowed by other cosmological constraints.

In addition to the \lya\ forest power spectrum measurement of
$\Delta^{2}(0.008 [$ km s$^{-1}]^{-1})=0.57^{+0.54}_{-0.27}$ (95\% CL) 
(we do not
use the \lya\ slope measurement, which has no significant
effect) at $z=2.5$,
we consider 6 other cosmological constraints.
The Hubble constant is measured to be $h=0.72 \pm 0.17$ (95\% CL),
where we have added statistical and systematic errors in quadrature
and doubled the $1\sigma$ errors
\cite{Madetal98}.    
The amplitude of the fluctuations is
determined by the COBE
detection of large angle anisotropies \cite{BunWhi97}; 
we ignore the 7\% measurement uncertainties on the temperature
fluctuations which are substantially smaller than the other uncertainties.
The abundance of galaxy clusters today
constrains models at the 8$h^{-1}$Mpc scale, where the
amplitude of fluctuations in top-hat spheres (at $z$=0) is
$\sigma_8 = 0.56 \Omega_m^{0.47}$ with 95\% CL of
$+20 \Omega_m^{0.2 \log \Omega_m}$ percent 
and 
$-18 \Omega_m^{0.2 \log \Omega_m}$ percent \cite{ViaLid98}.   

Galaxy surveys measure the shape of the power spectrum; 
we use the compilation of \cite{PeaDod94} and take
measurements only from the range 
$0.025 h$ Mpc$^{-1} < k < 0.25 h$ Mpc$^{-1}$
to avoid spurious survey volume effects on the large scale
and uncertainties in the nonlinear corrections on the small scale.  
For the galaxy survey data, we employ 
a $\Delta \chi^2$ statistic 
and take $\Delta \chi^2=4$ to represent the $95\%$ confidence
limits on the shape of the power spectrum. In carrying
this out, we assume that a linear ``bias'' is operating so that 
the matter power spectrum is related to the galaxy power spectrum
by a constant multiplicative factor (which we find as part of
our $\chi^2$ minimization).
We also employ nucleosynthesis constraints on the baryon density
of $\Omega_b h^2 = 0.019 \pm 0.0024$ (95\% CL) \cite{BurNolTruTur99}.
Finally, we place a lower limit on the age of the universe by assuming
that it must at least as old as the oldest globular clusters 
$(13.2 \pm 2.9)$Gyrs (95\% CL)
\cite{Caretal98}).

\begin{figure}[t]
\begin{center}
\leavevmode
\vskip -0.2truecm
\epsfxsize=3.4truein \epsfbox{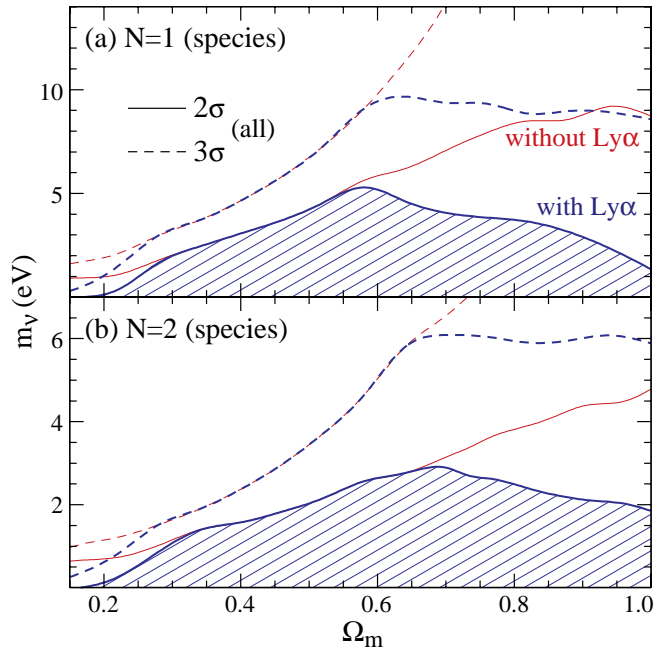}
\end{center}
\caption{Constraints on the neutrino mass (a) single massive
species (b) two (degenerate) mass species with and without the \lya\ 
constraint.}
\label{fig:neutrino_2}
\end{figure}

Given these constraints, we could construct a joint likelihood to find
the best fitting neutrino mass. We have decided to be more conservative,
however, and consider a model ruled out if it violates the
(95\%) confidence limits on any constraint taken individually.
With these ``$2\sigma$'' constraints on the parameter space, 
we use the analytic approximations of \cite{EisHu98} 
to explore the remaining space rapidly and  find the model that 
maximizes the neutrino mass as a function
of $\Omega_m$.
The result for one massive species is displayed
in Fig.~\ref{fig:neutrino_2}a.   
We also show the effect of omitting the 
\lya\ forest measurement.  
The measurement has a
powerful constraining effect at low and high $\Omega_m$ since the amount
of tilt required to match the cluster abundance and galaxy power
spectrum shape violates the upper and lower \lya\ forest
bounds respectively.  For no value of $\Omega_m$, can $m_\nu$
be greater than $5.5$eV.

To address the robustness of our upper limits, we show the effect
of scaling all errors by a factor of 1.5 to approximate
``$3\sigma$'' constraints in Fig.~\ref{fig:neutrino_2}.  
We also test for single point failures by dropping 
each constraint sequentially.   
Omission of either the age or cluster abundance constraint changes
the maximal neutrino mass to $\sim 7$eV. While dropping the   
 galaxy power spectrum constraint does not
increase the maximal mass substantially, it does weaken the bounds
by up to 2 eV for $\Omega_m \lesssim 0.5$. Omission of the
$h$ or $\Omega_b h^2$ constraint has essentially no effect.

In applying our constraints, we
assume that the universe is flat ($\Omega_m +\Omega_\Lambda=1$)
and that gravity waves do not contribute to the COBE normalization.  However,
assuming an open universe or gravity waves from power law inflation
does not change the $m_\nu$ limits significantly since the tilt
can be used to offset small changes in normalization.  
The simplest inflationary models can also predict a variation of the 
spectral index with scale. This is a small effect compared to the 
neutrino power suppression, so that we do not include it. Future CMB 
observations should address this point definitively.

We also show the results assuming 2 neutrino species with
identical masses in Fig.~\ref{fig:neutrino_2}b.  These limits are roughly
half the single species results since the
change in the growth rate is mainly governed by $\Omega_\nu$.
However dividing
the total mass into 2 species makes each species more relativistic
and enhances the suppression of the
power on scales relevant to the galaxy power spectrum
and cluster abundance constraints \cite{PriHolKlyCal95}.

\smallskip

{\it Future prospects.---} 
Additional  \lya\ forest data from quasar surveys such as
the Sloan Digital Sky Survey (SDSS \cite{SDSS}) have the potential 
to increase the precision of these mass constraints substantially.
Furthermore, we expect that the next generation of CMB satellites
will not only verify the existence of 
the underlying framework for structure formation,
which we currently assume, but also provide limits on the neutrino
mass itself. 

How precise must these \lya\
measurements be to improve on projected CMB limits on the neutrino mass?
To answer this question, we employ Fisher information matrix techniques
to approximate the joint covariance matrix.
For the \lya\ forest power spectrum 
measurement, the Fisher matrix is given by 
${\bf F}_{ij} = (\Delta P/P)^{-2} (\partial \ln P/\partial p_i)
				(\partial \ln P/\partial p_j)$.
We add this to the CMB Fisher matrix projected for the MAP and Planck
satellites including polarization information \cite{EisHuTeg98}.  
The variance of the optimal unbiased estimator 
of $p_i$ marginalized over the other parameters is $(F^{-1})_{ii}$.  

A fractional error of $\Delta P/P = 0.1$ (1$\sigma$) on the \lya\ power spectrum
would improve the MAP upper limit from
1.1 eV to 0.54 eV and the Planck limit from
0.51 eV to 0.29 eV both at 2$\sigma$.  Note that these represent limits
in a wider 10 parameter space including spatial curvature and gravity
waves.  These improvements would exceed those that can be 
achieved the SDSS galaxy survey \cite{EisHuTeg98}.
CMB polarization information here is critical for these 
improvements since they rely on an absolute normalization
of the power spectrum from degree scale anisotropies.  Polarization
information eliminates the degeneracy between the normalization and
the optical depth due to reionization.

Is 10\% precision in power achievable from \lya\ forest
measurements?  The constraint in this paper 
relies on data from $\sim 10$ full quasar spectra \cite{Pettinipk}.
The SDSS quasar survey \cite{SDSS}
 will yield spectra of roughly similar quality
(resolution $2.5$ \AA\ compared to $\sim1.5$ \AA\ for \cite{Pettinipk})
for $\sim 10^{5}$ quasars. The mean distance between sightlines
in the SDSS will be somewhat larger than the scale on which clustering
was measured by \cite{Pettinipk}, so that the decrease in 
statistical errors will not be too far off
the factor  $\sim (10^{5}/10)^{1/2} \sim 100$ which would be expected
if the sightlines were independent. A $1 \sigma$ statistical error of 
$\Delta P/P < 1\%$ should therefore be possible in the future,
so that systematic errors will become dominant.
Studying larger hydrodynamic simulations
should enable us to understand these systematic effects and refine our analysis
techniques. The vast size of the SDSS dataset will also 
enable us to pin down non-gravitational contributions to \lya\ forest 
clustering, for example by analyzing the evolution of the forest
with redshift. The signature of massive neutrinos, if they are present,
should therefore be obvious, even if $m_\nu$ is a fraction of an eV.

RACC acknowledges support from NASA Astrophysical
Theory Grant NAG5-3820. 
WH is
supported by NSF PHY-9513835, by a Sloan Fellowship and by
the WM Keck Foundation. RD is supported by  NASA Astrophysical
Theory Grant NAG5-7066 and by a Lyman Spitzer Fellowship.

 
\end{document}